\documentstyle[prl,aps,preprint]{revtex}

\begin{document}
\preprint{OUTP-95-62 S,~~
hep-th/9511134}
\title{\bf Logarithmic Operators and Hidden Continuous
Symmetry
in Critical Disordered
Models}
\author{J.-S. Caux, Ian  I. Kogan and A. M. Tsvelik}
\address{Department of Physics, University of Oxford, 1 Keble Road,\\
Oxford, OX1 3NP, UK}
\date{\today}
\maketitle

\begin{abstract}

 We study the model of (2 + 1)-dimensional relativistic fermions
in   a random non-Abelian gauge
potential at criticality. The exact solution shows that the
operator expansion contains a conserved current - a  generator
of a continuous symmetry. The presence of this operator  changes the
operator product expansion and gives rise to
logarithmic contributions to the correlation
functions at the critical point.
We calculate the distribution function
of the local density of states in this
model and find that it follows
the famous log-normal law.

\end{abstract}
\pacs{72.15.Rn}

\narrowtext

\section{Introduction}

 In this paper  we continue to study
the model of (2 + 1)-dimensional  massless Dirac
fermions interacting with a random static non-Abelian gauge
potential.
The Hamiltonian (or rather a generating
functional for the Green's functions) of this model is given by
\begin{eqnarray}
\mbox{i}\hat H - \epsilon_n\hat I =
\int \mbox{d}^2x\left(R^+_{\alpha}, L^+_{\alpha}\right)\left(
\begin{array}{cc}
(\partial_x - \mbox{i}\partial_y)\delta_{\alpha\beta} -
\mbox{i}A^+_{\alpha\beta}(x,y) & - \epsilon_n \\
- \epsilon_n & (\partial_x + \mbox{i}\partial_y) -
\mbox{i}A^-_{\alpha\beta}(x,y)
\end{array}\right)
\left(R_{\beta}, L_{\beta}\right)
\end{eqnarray}
where the random fields $A^a_{\alpha\beta}(x,y)~$($ a = \pm$)
 have a Gaussian distribution:
\begin{equation}
\langle A^a_{\alpha\beta}(\vec r_1)A^b_{\gamma\eta}(\vec r_2)\rangle =
{\cal A}  \delta(\vec r_1 - \vec
r_2)\delta_{a, -b}\delta_{\alpha\eta}\delta_{\beta\gamma}
\end{equation}
The fermionic  fields $R_{\alpha}, \: L_{\alpha}$ represent respectively the
right and the left moving components of the spinor field, and $\alpha$
takes the values $1,...,N$. This  model was introduced in
Ref. \cite{ners}
to describe normal
excitations in two
dimensional non-$s$-wave superconductors with disorder. In this
context  $N$ denotes
the number of nodes of the order parameter on the Fermi surface.

 Since  the disorder is time-independent we consider
the Fourier components of the fermionic fields with
different frequencies separately. This reduces the dimensionality of
the problem, making it two dimensional.  The model (1) is exactly
solvable in the subspace   $\epsilon = 0$, where it is
critical.  The
presence of the superconducting order parameter fixes the chemical
potential and thus insures  that
the disorder is diagonal in
chirality. This is essential for the criticality  at $\epsilon
= 0$. At the critical point,  one can  apply  the
methods of conformal field theory and calculate scaling dimensions of
the fields and of their
multi-point  correlation functions.  This gives us a rare opportunity
to obtain  nonperturbative results for a
non-trivial random theory.
We hope  that a study of this exact solution will  give an insight into
general properties of random systems.

  The averaging over the disorder can be done either
through  the replica trick
\cite{tsv} or using the supersymmetric  approach
\cite{wen},\cite{gade}. In this paper we shall
mostly  use the replica approach with which
we are  more familiar.
To demonstrate  that the two approaches
are equivalent we shall (i) compare the conformal
dimensions of the primary fields and (ii)
demonstrate that both representations  give the same
conformal blocks for the four-point correlation function.

 As in the standard localization theory (see for example
\cite{efetov}), one can integrate out the fast degrees of freedom and
derive an effective action for the slow ones in the form of a
sigma model. This program was carried out in Ref.\cite{ners}. The
resulting sigma model has the following action:
\begin{eqnarray}
S = S_0 + M\epsilon_n \int\mbox{d}^2x \mbox{Tr}(Q + Q^+)
\end{eqnarray}
 where the $S_0$ action  contains the Wess-Zumino term:
\begin{eqnarray}
S_0 &&= \frac{Nr}{2}\int \mbox{d}^2x (\partial_{\mu}\Phi)^2 + NW[SU(r);
g]
\nonumber\\\vspace{.5cm}
W[SU(2r); g] &&= \frac{1}{16\pi}\int \mbox{d}^2x\left[
\mbox{Tr}(\partial_{\mu}g^+\partial_{\mu}g) +
\frac{2}{3}\int_0^{\infty}\mbox{d}\xi
\epsilon_{abc}
\mbox{Tr}(g^+\partial_{a}g g^+\partial_b g g^+\partial_c g)\right] \label{wzw}
\end{eqnarray}
where $Q$ is
the $r\times r$ matrix $Q = g\exp[\mbox{i}\sqrt{4\pi}\Phi]$, and where $g$
belongs to the SU(r) group.  $\Phi$ is a  real scalar field defined
on the  cirle with circumference $\sqrt\pi$.
The quantity $M$ is the energy
scale introduced by the disorder: $M \sim \exp[ - 2\pi/N {\cal A}]$;
it  marks the crossover from the bare density of states (DOS)
$\rho(\epsilon) \sim |\epsilon|$ to  the renormalized
DOS $\rho(\epsilon) \sim |\epsilon|^{\nu}$ ($\nu = (2N^2 - 1)^{-1}$).
$M$ serves as the ultraviolet cut-off for the sigma
model  (\ref{wzw}).

 It is well known that in  conventional
localization theories the {\it average} DOS is not affected by the
disorder (its higher moments are, however).
It is not the case for  the model (1), where the DOS is directly
proportional  to the order parameter. As it was shown in \cite{ners},
the local DOS is given by
\begin{equation}
\rho(\epsilon, x) = \frac{M}{r}\mbox{Tr}[Q(x) +
Q^+(x)] \label{rho}
\end{equation}
This means that even the average DOS is strongly renormalized.
At $\epsilon$ = 0 the sigma model (\ref{wzw}) is critical and
$\langle \rho(0, x)\rangle = 0$, as it might be expected in a critical
theory in two dimensions \cite{remark}. In three dimensions a finite DOS
emerges at $\epsilon = 0$
\cite{gorkov}. In our previous publications we interpreted this
effect as a manifestation of
violation of some continuous symmetry present in the
theory (\cite{ners}, \cite{tsv}). However, the meaning of this
symmetry has remained obscure.  In this paper we identify  the
operators which  generate this symmetry and derive  their algebra.

 The paper is organized as follows. In Section II we discuss general
properties of the model at criticality and derive the expression for
the conformal dimensions of its primary fields. In Section III we derive
the expression for the four-point correlation function of local DOS.
It turns out that this correlation function contains logarithmic
singularities and therefore  a fusion of local DOS generates
operators with unusual
 properties - the  so-called logarithmic operators.
 In Section IV we develop a general theory of such operators and
demonstrate that the appearance of such operators  always implies the
presence of some  continuous
symmetry. It may well be that this
symmetry is present in all critical models with disorder.
In Section V we show how  these general  results hold
for the model (4). In Section VI
we  study  correlation functions  in the  model
deformed  by the logarithmic operators away from
criticality. It
turns  out that even in the case of a marginal deformation the
 correlation  functions have logarithmic corrections.   In Section VII
we derive the log-normal  distribution function of the local DOS.
The paper contains a conclusion and an appendix
where the conformal blocks of replica and supersymmetric
theories are compared.

\section{Conformal dimensions}

 The  WZNW model has been
well studied at  finite $r$. There is an extensive
literature on the subject, but we particularly recommend
the original publication by Knizhnik and Zamolodchikov\cite{knizh}.
 These  authors derived  explicit expressions for
the four-point correlation functions of primary fields which we are going to
exploit. In our calculational procedure
we follow the
general principle: when calculating
any $n$-point correlation function $F_{r}(1, 2, ...n)$
$r$ is treated
as an arbitrary  number on all intermediate steps
of the calculations until the final expression is
obtained. We define the replica limit as follows:
\begin{equation}
F(1,2, ... n) = \lim_{r \rightarrow 0}\frac{2N}{r}F_{r}(1, 2, ...n)
\label{eq:def}
\end{equation}
The reason for the introduction of  the extra factor  $2N$ will be
discussed later.

 Let us study the correlation functions of the $Q, Q^+$-fields. The problem
of indices is simplified
by the fact that the $Q_{pq}$ matrices are slow parts of the
operators
\begin{equation}
Q_{pq} \sim \sum_{\alpha = 1}^N R^+_{\alpha,p}L_{\alpha,q}
\end{equation}

 From this fact one can derive  a simple recipe for the index structure of
$n-$point correlation functions: it is the same as for the
$n$-point function of the
$\sum_{\alpha}R^+_{\alpha,p}L_{\alpha,q}$-fields
in  the theory of
massless free fermions. The simplest example
is the 2-point function \cite{tsv}, \cite{wen}:
\begin{eqnarray}
\langle Q_{p_1q_1}(z,\bar z)Q^+_{q_2p_2}(0,0)\rangle =
\delta_{q_1q_2}\delta_{p_1p_2}\frac{1}{(M|z|)^{2/N^2}} \label{eq:corr}
\end{eqnarray}
where
$1/2N^2$ is the conformal dimension of the composite
operator $Q$ given by the sum of the dimensions of the bosonic
exponent $\exp[\mbox{i}\sqrt{4\pi}\Phi]$ and of the operator
field  $g_{pr}$ from the
fundamental representation of the SU(r) group:
\begin{equation}
\Delta = \lim_{r \rightarrow 0}\left[\frac{1}{2rN} + \frac{(r - 1/r)}{2(N +
r)}\right] = \frac{1}{2N^2}  \label{eq:dimen}
\end{equation}
In the replica limit we get from Eq.(\ref{eq:corr})
\begin{equation}
G(z, \bar z) \equiv \lim_{r\rightarrow 0}(2N/r)
\langle\mbox{Tr}[Q(z, \bar z)]
\mbox{Tr}[Q^+(0,0)]\rangle = (M|z|)^{- 2/N^2} \label{eq:corr1}
\end{equation}

 All other operators are generated by fusion of the fundamental
fields Tr$Q$ and Tr$Q^+$.
The corresponding primary fields are composite fields of
bosonic exponents and Wess-Zumino tensors belonging  to irreducible
representations of the SU(r) group. These representations are
classified by Young tableaus which can be represented by a string
of numbers $f_1 > f_2 > ... > f_{r} \geq 0$. Only representations
with $f_l \leq N - 1$ are generated \cite{knizh}.
The corresponding
conformal dimensions are given by the expressions
\begin{eqnarray}
\Delta_f &&= \frac{C_f}{N + r} + \frac{f^2}{2rN}\nonumber\\
C_f &&= \frac{1}{2}\sum_l[f_l^2 + (r + 1 - 2l)f_l] - \frac{f^2}{2r}
\end{eqnarray}
where $f = \sum_lf_l$. In the replica limit we get
\begin{eqnarray}
\Delta_f = \frac{f^2}{2N^2} + \frac{1}{2N}\sum_l[f_l^2 - (2l - 1)f_l]
\label{dims}
\end{eqnarray}
This expression coincides with the conformal dimensions obtained by
the supersymmetric approach after rows and columns in the Young
tableau are interchanged. For instance,
for the representation $(\overbrace{1,1, ... 1}^m, 0,..)$ we reproduce
the expression obtained in \cite{wen}(see Eq.(4.48b)there) for the
$(m, 0, ...0)$ representation:
\begin{equation}
\Delta_m = \frac{m}{2N}\left[1 - \frac{(N - 1)m}{N}\right] \label{deltas}
\end{equation}

\section{Four point correlation function of the order parameter
fields}

 Let us now study the four point correlation function of the $Q,~ Q^+$
fields. The index structure is the same for all $r$\cite{knizh}:
\begin{eqnarray}
\langle Q_{p_1q_1}(z_1,\bar z_1)Q^+_{q_2p_2}(z_2,\bar z_2)&&Q_{p_3q_3}
(z_3,\bar z_3)Q^+_{q_4p_4}(z_4, \bar z_4)\rangle \nonumber\\
&&= M^{-4/N^2}\left[\frac{|z_{13}z_{24}|}{|z_{12}z_{14}z_{23}z_{34}|}
\right]^{2/N^2}(W
+ \tilde W),\\
\tilde W &&=
[\delta_{p_1p_2}\delta_{p_3p_4}\delta_{q_1q_2}\delta_{q_3q_4}
W_{11}(x,\bar
x) + \delta_{p_1p_3}\delta_{p_2p_4}\delta_{q_1q_3}\delta_{q_2q_4}
W_{22}(x,\bar x)],\\
W &&= [\delta_{p_1p_2}\delta_{p_3p_4}\delta_{q_1q_3}\delta_{q_2q_4}
W_{12}(x,\bar x) + \delta_{p_1p_3}\delta_{p_2p_4}\delta_{q_1q_2}
\delta_{q_3q_4}W_{21}(x,\bar x)]
\end{eqnarray}
where
\begin{equation}
x = \frac{z_{12}z_{34}}{z_{13}z_{24}}, \: \bar x = \frac{\bar z_{12}\bar
z_{34}}{\bar z_{13}\bar z_{24}} \label{z}
\end{equation}
Here the functions  $W_{AB}(x,\bar x), \: (A,B = 1,2)$ satisfy
linear differential equations (the Knizhnik-Zamolodchikov equations)
which we shall discuss later in detail.
Now note
that in our theory we shall deal only with correlation functions of
Tr$Q$, Tr$Q^+$ (for simplicity we do not consider transport phenomena
which would need introduction of advanced and retarded correlation
functions).
Since all correlation functions must be
proportional to $r$, only the $W$ term (that is the term with
all indices equal) survives
in the replica limit,
the $\tilde W$ term being proportional to $r^2$.
Therefore we have
\begin{eqnarray}
(2N/r)\langle\mbox{Tr}
&&Q(1)\mbox{Tr}Q^+(2)\mbox{Tr}Q(3)\mbox{Tr} Q^+(4)\rangle
\nonumber\\
=&& 2N M^{-4/N^2}\left[\frac{|z_{13}z_{24}|}{|z_{12}z_{34}z_{23}z_{3
4}|}\right]^{2/N^2}[W_{12}(x,\bar
x) + W_{21}(x,\bar x)] \label{four1}
\end{eqnarray}

 The functions $W_{AB}(x, \bar x) = U^{pq}W^{(p)}_A(x)W^{(q)}_B(\bar x)$
are composed of linearly independent solutions of the
Knizhnik-Zamolodchikov equation, written $W^{(p)}_A(x)$ and $ \: W^{(q)}_B(\bar
x)$
(conformal blocks). In the replica limit these equations
have  the following form:
\begin{eqnarray}
Nx\frac{d W_1}{dx} = - W_2, ~~~~~~~~~~~~~
 N(1 - x)\frac{d W_2}{dx} = W_1
\end{eqnarray}
Thus for the function $W_1$ we get the following hypergeometric
equation:
\begin{equation}
N^2\frac{d}{dx}\left(x\frac{d W_1}{dx}\right) + \frac{W_1}{1 - x} = 0
\label{equ}
\end{equation}
Here we encounter a problem.
A hypergeometric equation has always two  linearly independent
solutions, normally expressed in terms of powers and hypergeometric functions.
Usually there are three sets of  solutions defined in the
vicinity  of  $x =
0$, 1 and $\infty$ respectively. These pairs  of solutions are related
to each other via simple  transformation rules (see, for example
\cite{abram}). Eq.(\ref{equ}) is an exlusion: in the vicinity of
$x = 0$  one of the  solutions contains a
logarithmic singularity and cannot be expressed in terms of
hypergeometric functions (this second solution was
overlooked in the previous publication of one of the authors \cite{tsv}):
\begin{eqnarray}
W_1^{(0)}(x) &&= F(1/N, - 1/N, 1, x), \nonumber\\
\: W_1^{(1)}(x) &&= \ln x
W_1^{(0)}(x) + H_1(x)\nonumber\\
NW_2^{(0)}(x)  &&= x F(1 + 1/N, 1
- 1/N, 2, x), \nonumber\\
 NW_2^{(1)}(x) &&= \ln x W_2^{(0)}(x) - N^2 + H_2(x) \label{zero}
\end{eqnarray}
where $H_{1,2}(x)$ are functions that are regular at $x = 0$:
\begin{eqnarray}
H_1(x) = \sum_{n = 1}^{\infty}&&\frac{x^n(1/N)_n(-1/N)_n}{(n!)^2}
  [\psi(1/N + n) - \psi(1/N) + \nonumber\\
 &&+ \psi(- 1/N + n) - \psi(- 1/N) - 2\psi(n + 1) + 2\psi(1)] \nonumber\\
H_2(x) = x\sum_{n =
0}^{\infty}&&\frac{x^n(1 + 1/N)_n(1 - 1/N)_n}{n!(n+1)!}
[\psi(1 + 1/N + n) - \psi(1 + 1/N) + \nonumber\\
 &&+ \psi(1 - 1/N + n) - \psi(1 - 1/N)
 - \psi(n + 1) + \psi(1) - \psi(n + 2) + \psi(2)]
\end{eqnarray}
where $(a)_n = \Gamma(a + n)/\Gamma(a)$.
 Only in the vicinity of $x = \infty$ are the solutions still
hypergeometric functions (for $N \neq 2$).

At $|x| << 1$ we have
\begin{eqnarray}
W_1^{(0)}(x) = 1 + O(x), ~~~~~&&~~~~~  W_1^{(1)}(x) = \ln x [1 + O(x)],
\nonumber\\
NW_2^{(0)}(x)  = x + O(x^2), ~~~~~&&~~~~~  NW_2^{(1)}(x) = - N^2 + x\ln x[1 +
O(x)]
\end{eqnarray}

Now we have to choose the matrix $U^{pq}$ in such a way that the
resulting expression for the four point correlation function be
a uniquely defined function in the complex plane of $x$. It also must be
invariant under the permutation of points 1 and 3 (2 and 4) which
means the invariance under $x \rightarrow 1 - x, \: \bar x \rightarrow
1 - \bar x$ (crossing symmetry). These properties are achieved when
\begin{eqnarray}
U^{(01)} = U^{(10)}, \: U^{(11)} = 0, \: U^{(00)} = hU^{(01)}
\end{eqnarray}
 To find $h$, we first note that the solutions to the Knizhnik-Zamolodchikov
equations obey the monodromy
properties
\begin{eqnarray}
W_1^{(0)}(1-x) = a_i W_2^{(i)}(x)\nonumber\\
W_2^{(0)}(1-x) = b_i W_1^{(i)}(x)
\end{eqnarray}
where
\begin{eqnarray}
a_0 &&= a_1 [\psi(1/N) + \psi(-1/N) - \psi(2) - \psi(1)]\nonumber\\
a_1 &&= b_1 = \frac{N}{\Gamma(1/N) \Gamma(-1/N)}\nonumber\\
b_0 &&= a_1 [\psi(1/N) + \psi(-1/N) - 2\psi(1)]
\end{eqnarray}
 Using the crossing symmetry, we then find that
\begin{equation}
h = \frac{a_0 b_1 + a_1 b_0}{a_1 b_1} = 1/2
\end{equation}
 Thus we get
\begin{eqnarray}
G(1,2,3,4)
 =&&
- \frac{1}{2N}
\left[\frac{|z_{13}z_{24}|M^{-2}}{|z_{12}z_{14}z_{23}z_{34}|}\right]^{2/N^2}
\nonumber\\
&&\times[W_1^{(0)}(x)W_2^{(1)}(\bar
x) + W_1^{(1)}(x)W_2^{(0)}(\bar
x) + \frac{1}{2}W_1^{(0)}(x)W_2^{(0)}(\bar
x) + (x \rightarrow \bar x)] \label{four}
\end{eqnarray}
Here we choose $U^{(01)} = - 1/4N^2$ for normalization.

 In order to derive  the operator algebra of the model we consider
various limits of this formula. In the limit $z_{43} = \epsilon
\rightarrow 0$ we get
\begin{equation}
\langle [QQ^+(3)]Q(1)Q^+(2)\rangle = \frac{1}{|\epsilon
z_{12}|^{2/N^2}}\left[1 - \frac{1}{N^2}\left(\frac{\epsilon
z_{12}}{z_{13}z_{23}} +
c.c\right)\ln\left(|\epsilon||\frac{z_{12}}{z_{13}z_{23}}|\right) + ...\right]
\label{three}
\end{equation}
 From now on we shall use $Q$ without subscripts instead of Tr$Q$
assuming that the replica limit has been taken. We shall also put $M =
1$.

 The three-point correlation function (\ref{three}) is very unusual
from the conformal field theory point of view because it
contains logarithms. Therefore we pause to consider general properties
of logarithmic operators.

\section{General properties of  logarithmic operators}

 So far correlation  functions with logarithms at criticality
have been obtained in the WZNW model on the supergroup $GL(1,1)$
 \cite{rs}, in   the C = - 2  model
\cite{gurarie} and in gravitationally dressed  CFT \cite{bk}.
 It was first pointed out by Gurarie in  Ref. \cite{gurarie}
  that the appearence of logarithms in correlation  functions is
due to the presence of special operators, whose operator product
expansions (OPE's) display
logarithmic short-distance singularities. These logarithmic
operators have conformal dimensions degenerate with those of the
usual primary operators, and it is this degeneracy that is at the
origin of the logarithms (cf. our discussion of the degenerate
hypergeometric equation in the previous Section). As a result of this
degeneracy one can
no longer completely diagonalize the Virasoro operator $L_0$, and the
new operators together with the standard ones form the basis of the
Jordan-cell for $L_0$.  In order to get a better insight in the
situation, we shall consider the simplest example, which was mentioned
  in \cite{bk}, namely,  the Liouville model with the action
\begin{equation}
S = \frac{1}{8\pi} \int \mbox{d}^2\xi \sqrt{g(\xi)}\left[
\partial_{\mu}\phi(\xi) \partial^{\mu}\phi(\xi) + Q R^{(2)}(\xi)
\phi(\xi)
 \right]
\label{Liou}
\end{equation}
($R^{(2)}$ is the Riemann curvature on a two-dimensional surface - the
world  sheet)  with the stress-energy  tensor
$$
T = -\frac{1}{2} \partial_{z} \phi \partial_{z} \phi +
 \frac{Q}{2} \partial_{z}^2 \phi
$$
 and the  central charge
\begin{equation}
{\cal C} = 1 + 3 Q^2 \label{central}
\end{equation}
 The primary field
 $\exp(\alpha \phi)$ has a dimension
\begin{equation}
\Delta_{\alpha} =
 \alpha(Q-\alpha)/2 \label{Eq}
\end{equation}
 This means that there are two operators
 with the same dimension $\Delta_{\alpha}$, namely
 $V_{\pm} = \exp( \alpha_{\pm} \phi)$, where
$$
\alpha_{\pm} = \frac{Q}{2} \pm \frac{1}{2}\sqrt{Q^2 -
8\Delta_{\alpha}}
$$
If  $Q^2 = 8\Delta_{\alpha}$,
i.e.   when $\alpha = Q/2$,   there is a degeneracy $\alpha_{+} =
 \alpha_{-}$ and instead of two exponential primary fields we have
 only  one exponent $C = \exp(\frac{1}{2}Q\phi)$ and the new operator
 $ D = \phi \exp(\frac{1}{2}Q\phi)$ with the same dimension
 $\Delta = Q^2/8$. The latter  field is sometimes called
 the puncture operator. It was discussed in \cite{polch} in the context of the
 Liouville  gravity,   when  the action (\ref{Liou}) describes the
 gravitational (Liouville) sector of a (non)critical string in the
conformal  gauge.

 It is easy to get the OPE of the stress-energy
 tensor $T$ with these fields. After simple calculations we find
\begin{eqnarray}
T(z) C(0) &&=  {\Delta \over z^2} C(0)+ {1\over z} \partial_z C(0) +
...
\nonumber\\
T(z)  D(0) &&=  {\Delta \over z^2} D(0)+{1\over z^2}
 C(0)+{1\over z} \partial_z D(0) + ...
\label{JOPE}
\end{eqnarray}
 where the dimension of the fields $C$ and $D$ is $\Delta = Q^2/8$
 and the normalization of the field $D$ was defined as
 $ D =(2/Q)\phi \exp(\frac{1}{2}Q \phi)$.

 It is easy to  see indeed that there is  a  mixing  between $C$ and $D$
 and that the   Virasoro operator $L_0$  which is defined through the
 Laurent expansion $T(z) = \sum_{n} L_{n} z^{-n-2}$ is not diagonal
\begin{eqnarray}
L_{0}|C> = \Delta |C>, ~~~~~~ L_{0}|D> = \Delta |D> + |C> \label{example}
\end{eqnarray}
 Let us also note that usually one can think about
factorization of the primary
field  into the product of  chiral left and right operators using the
 decomposition $\phi(z, \bar{z})  = \phi_L (z) + \phi_R (\bar{z})$,
leading to $\exp[\alpha \phi] = \exp[\alpha  \phi_L (z)] \times
 \exp[\alpha \phi_R (\bar{z})]$. For a logarithmic operator
 we have
$$
\phi ~\exp[\alpha \phi] = \phi_L (z) \exp[\alpha  \phi_L (z)] \times
 \exp[\alpha \phi_R (\bar{z})] + \phi_R (\bar{z})
\exp[\alpha  \phi_L (z)] \times  \exp[\alpha \phi_R (\bar{z})]
$$
and thus the logarithmic operator is the sum of left and right
operators each of which  can be factorized.

 This simple example illustrates a  quite general property of all
theories with logarithmic operators and OPE (\ref{JOPE}) is valid
 in all these theories. One can also obtain some general information
 about two- and three-point correlation functions with operators $C$
 and $D$ starting from the four-point correlation function \cite{gurarie}
\begin{equation}
\langle A(z_1)B(z_2) A(z_3) B(z_4)\rangle =
\frac{1}{(z_1-z_3)^{2\Delta_{A}}(z_2-z_4)^{2\Delta_{B}}}
[x(1-x)]^{\Delta_C - \Delta_{A} - \Delta_{B}} F(x)
\end{equation}
  where $x$ is defined by Eq.(\ref{z}),  and where we have extracted
 the factor $[x(1-x)]^{\Delta_C - \Delta_{A} - \Delta_{B}}$ to
 make  $F(x)$ finite at  $x \rightarrow 0$ or
 $x \rightarrow 1$ in the  ordinary  case. In the case of logarithmic
 operators one will get  $F(x) = (d + c \ln x + o(x))$  at small $x$.
 To reproduce the logarithmic singularity at $x=0$ after the fusion
 of $A(z_1)$ and $B(z_2)$ one has to postulate the following OPE
 (we restrict ourselves to the  chiral sector):
\begin{equation}
A(z_1)~ B(z_2) = (z_1-z_2)^{\Delta_C - \Delta_A -\Delta_B}
\left[D + C \ln(z_1-z_2) +..\right]
\label{AB}
\end{equation}
Taking the limit $z_1 \rightarrow z_2$ one immediately gets
 from the four-point correlation function the following three-point
 correlation functions:
\begin{eqnarray}
\langle C(z_1)  A(z_3) B(z_4)\rangle  &&=
\frac{c}{z_{13}^{\Delta_{A}+\Delta_C-\Delta_B}
 z_{14}^{\Delta_{B}+\Delta_C-\Delta_{A}}z_{34}^{\Delta_{A} +
 \Delta_{B} - \Delta_C}}~  \nonumber \\
\langle D(z_1)  A(z_3) B(z_4)\rangle  &&=
\frac{1}{z_{13}^{\Delta_{A}+\Delta_C-\Delta_B}
 z_{14}^{\Delta_{B}+\Delta_C-\Delta_{A}}z_{34}^{\Delta_{A} +
 \Delta_{B} - \Delta_C}}~\left(c \ln \frac{z_3-z_4}{(z_1-z_3)(z_1 - z_4)}
 + d \right)
\label{CAB}
\end{eqnarray}

Now  let us consider the  $ A(z_3)$ and $ B(z_4)$ fusion  which
 after insertion of (\ref{AB}) into (\ref{CAB}) will lead to the
 following two-point  correlation functions:

\begin{eqnarray}
\langle C(x) D(y)\rangle &&=
\langle C(y) D(x) \rangle  = \frac{c}{(x-y)^{2\Delta_C }}\nonumber \\
\langle D(x) D(y)\rangle &&=
 \frac{1}{(x-y)^{2\Delta_C}} \left(-2c\ln(x-y) + d\right)
\nonumber \\
\langle C(x) C(y)\rangle  &&= 0
\label{CC}
\end{eqnarray}

The first equation imposes a strong constraint on the
  dimensions $\Delta_C$ of  logarithmic
 operators, namely  that the  dimension $\Delta_C$ must be {\it an integer}.
 To  prove  it let us note that we  have
   $\langle C(x)D(y)\rangle =\langle C(y)D(x)\rangle $,
 which means that
 the correlation function is invariant under the permutation of
$x$  and $y$, which means that $(-1)^{2\Delta_C} = 1$ so $\Delta_C =n$.
In case of noninteger $\Delta_C$ one must have  the
structure constant $c = 0$  and only the  operator $D$ will survive in
OPE (\ref{AB}),
however in this case  it will be an ordinary, nonlogarithmic operator.

  This new result  about dimensions  of  logarithmic
 operators means  that for any
 logarithmic operator we have a hidden continous symmetry. This
symmetry is generated by  the conserved
holomorphic (or antiholomorphic) current $C(z)$.
This current is a  symmetric
tensor of rank $\Delta_{C}$, which is
 a usual vector  current if $\Delta_{C} = 1$. In the next Section we
shall demonstrate the existence of such a  conserved  vector
current in the  model with  disorder that we are considering.
 Let us note that we have
 also proved that there is no central extension in the corresponding
current algebra, or in other words there is no anomalous Schwinger
term  in the current-current commutator. This is the direct
consequence of the triviality of
the correlation function $\langle C(z)C(0)\rangle = 0$.

\section{Operator product expansions in the model (4)}

 In this Section we demonstrate how the general theory just discussed
applies to the model (4). For this end we study OPE's in this  theory
after the replica
limit has been taken. In doing so we assume that the replica limit can be
described by some quantum field theory. Here we encounter  a
certain ambiguity, namely, that  we can define  the  correlation
functions with an arbitrary prefactor. It turns out that this
prefactor is fixed by the requirement of
self-consistency of OPE's. The latter  is achieved when one
uses the definition
(\ref{eq:def}). This explains the necessity  of the factor
$2N$ in Eq.(\ref{eq:def}).

 We suggest
the following OPE:
 \begin{eqnarray}
Q(z)Q^+(0)
= |z|^{-2/N^2}\left\{I - z\left[D(0) +
C(0)\ln|z|^2\right]  - \bar z\left[\bar D(0) +
\bar C(0)\ln|z|^2\right] ...\right\} \label{ope}
\end{eqnarray}
where $D, C$ and $\bar D, \bar C$
are some new operators whose correlation functions are to be  found.
Notice that in the conventional
WZNW theory this operator expansion would
contain the unit operator and the operator in the adjoint
representation. However, the latter one has the conformal dimensions
which vanish in the replica limit:
\begin{equation}
\Delta_{ad} = \bar\Delta_{ad} = \frac{c_v}{c_v + N} = \frac{r}{r +
N} \rightarrow 0
\end{equation}
Therefore we have here a situation described in the previous Section:
the conformal dimensions  of descendants of the unity operator become
degenerate with the dimensions of
descendants of some other primary field (the adjoint operator)
which gives rise to logarithms.

Substituting (\ref{ope}) into Eq.(\ref{three}) we get
\begin{eqnarray}
\langle Q(1)Q^+(2)C(3)\rangle &&=
 \frac{1}{2N^2}|z_{12}|^{-2/N^2}\frac{z_{12}}{z_{13}z_{23}}\nonumber\\
\langle Q(1)Q^+(2)D(3)\rangle &&=
N^{-2}|z_{12}|^{-2/N^2}\frac{z_{12}}{z_{13}z_{23}}
\ln|\frac{z_{12}}{z_{13}z_{23}}|
\label{qc}
\end{eqnarray}
Setting $z_{12} = \epsilon$ in these equations and using the OPE
(\ref{ope}) we get the following set of two-point correlation
functions:
\begin{eqnarray}
\langle D(1)C(2)\rangle &&= - \frac{1}{2N^2{z_{12}}^2}, \nonumber\\
 \langle C(1)C(2)\rangle &&= 0, \nonumber\\
\langle D(1)D(2)\rangle &&=  \frac{2 \ln|z_{12}|}{N^2{z_{12}}^2} \label{cddd}
\end{eqnarray}
There are similar expressions for $\bar C, \: \bar D$-operators with
$z$ being substituted for $\bar z$. Correlators of $C, \: D$ and $\bar C,
\: \bar D$ are equal to zero.

Setting $z_{31} = \epsilon$ in Eqs.(\ref{qc}) we deduce the following
OPE:
\begin{eqnarray}
C(z)Q(0) &&=  \frac{1}{2N^2z}Q(0) + ... , \nonumber\\
 D(z, \bar z)Q(0) &&= -
\frac{1}{N^2z}\ln|z|Q(0) + ... \label{cdq}
\end{eqnarray}
with the same equations for $Q^+$, except for a change of sign.
These OPE's and the fact that
$C(z)$ does not depend on $\bar z$ ($\bar\partial\langle
C(z)D(0)\rangle
 = 0$), enable us to
identify $C, \bar C$ as
generators of a  continuous symmetry. It is this symmetry
 which is associated with the order parameter $\rho$.

 Conformal field theories are  characterized by their symmetry
group and a number  ${\cal C}$
called `conformal charge'. Formally ${\cal C}$ is a coefficient in the pair
correlation function of stress-energy tensor operators.
A physical meaning of ${\cal C}$ becomes clear when  we recall that a
theory with an integer conformal charge
${\cal C} = k$ is equivalent to the theory with $k$ species of free
bosonic
 fields.
Thus ${\cal C}$ in unitary theories counts an effective number of degrees of
freedom. The central charge of our theory
is the
sum of central charges
of the free bosonic field (${\cal C} = 1$) and  the
WZNW model on the SU(r) group:
\begin{equation}
{\cal C} = 1 + \frac{N(r^2 - 1)}{N + r} = \frac{r}{N} + O(r^2)
\end{equation}
Thus the resulting
central charge vanishes, as it must be; however,
according to the definition of the replica limit
(\ref{eq:def})
the  physical correlation function
of the stress-energy tensors remains finite:
\begin{equation}
\langle T(z)T(0)\rangle = \lim_{r\rightarrow 0}\frac{2N{\cal C}_r}{r}
\frac{1}{2z^4} =
\frac{1}{z^4} \label{c}
\end{equation}
Superficially this looks like  the effective central charge  ${\cal C}_{eff} =
2$. However, ${\cal C}_{eff}$ does not appear in  the fusion rules of the
stress-energy tensor components inside of correlation functions with
matter fields, where we have
\begin{equation}
T(z)T(\xi) = \frac{2}{(z - \xi)^2}T(\xi) + \frac{1}{z -
\xi}\partial_{\xi}T(\xi) + ...
\end{equation}
As we have mentioned above, the numerical coefficient in (\ref{c})
is fixed by the
self-consistency requirements of OPE.

 Applying twice the OPE (\ref{ope}) to the four-point correlation
function and using the Ward identities for $Q$ fields,
we get the following set of identities:
\begin{eqnarray}
\langle T(z)C(1)D(2) \rangle &&= \sum_{j=1,2}\left\{ \frac{1}{(z - z_j)^2} +
\frac{1}{z - z_j} \partial_j \right\} \langle C(1)D(2) \rangle \nonumber\\
\langle \bar T(z)C(1)D(2) \rangle &&= \sum_{j=1,2} \frac{1}{\bar z - \bar z_j}
\bar \partial_j \langle C(1)D(2) \rangle = 0 \nonumber\\
 \langle T(z)D(1)D(2) \rangle &&= \sum_{j=1,2}\left\{ \frac{1}{(z - z_j)^2} +
\frac{1}{z - z_j} \partial_j \right\} \langle D(1)D(2) \rangle + \sum_{j=1,2}
 \frac{1}{z - z_j} \langle C(1)D(2) \rangle \nonumber\\
\langle \bar T(z)D(1)D(2) \rangle &&= \sum_{j=1,2} \left\{ \frac{1}{\bar
z -
 \bar
z_j} \bar \partial_j \right\} \langle D(1)D(2) \rangle + \sum_{j=1,2}
 \frac {1}{(\bar z - \bar z_j)^2} \langle C(1)D(2) \rangle
\end{eqnarray}
 We can then substitute the two-point correlation functions to get

\begin{eqnarray}
\langle T(z)C(1)D(2)\rangle &&= - \frac{1}{2N^2}\frac{1}{(z - z_1)^2(z -
z_2)^2}\nonumber\\
\langle T(z)D(1)D(2)\rangle &&= \frac{2}{N^2}\frac{1}{(z - z_1)^2(z -
z_2)^2}(\ln|z_{12}| - 1/4)\nonumber\\
\langle T(z)\bar D(1)\bar D(2)\rangle &&= -
\frac{1}{2N^2}\frac{z^2_{12}}{(z -
z_1)^2(z -
z_2)^2\bar z_{12}^2}
\end{eqnarray}
Taking into account Eq.(\ref{c}) and Eqs.(\ref{cddd})
we conclude that these  expressions
are  compatible with the following OPE:
\begin{eqnarray}
C(z)D(\xi, \bar\xi) &&= -
\frac{1}{2N^2}\left[\frac{1}{(z - \xi)^2} + T(\xi) + ...\right]\label{CD}\\
D(z,\bar z)D(\xi, \bar\xi) &&= \frac{2}{N^2}\left[\frac{\ln|z - \xi|}{(z
- \xi)^2} +
(\ln|z - \xi| - 1/4)T(\xi) - \frac{(\bar z - \bar\xi)^2}{4(z
- \xi)^2}\bar T(\bar\xi) + ...\right] \label{DD}
\end{eqnarray}
and
\begin{eqnarray}
T(z)C(\xi) &&= \frac{C(\xi)}{(z - \xi)^2} +
\frac{\partial_{\xi}C(\xi)}{(z - \xi)} + ...\\
T(z)D(\xi, \bar\xi) &&= \frac{D(\xi, \bar\xi)}{(z - \xi)^2} + \frac{C(\xi)}{(z
- \xi)^2} +
\frac{\partial_{\xi}D(\xi, \bar\xi)}{(z - \xi)} + ...\\
\bar T(\bar z)D(\xi, \bar\xi) &&= \frac{C(\xi)}{(\bar z - \bar\xi)^2} +
\frac{\partial_{\bar\xi}D(\xi, \bar\xi)}{(\bar z - \bar\xi)} + ...\\
\end{eqnarray}

 From the OPE's (\ref{cdq}, \ref{CD}), the Ward identity for the
stress-energy tensor and primary fields  and the
Knizhnik-Zamolodchikov
equation
(\cite{knizh}) we derive  the following Ward identity:
\begin{eqnarray}
&&\langle C(z_1)D(z_2, \bar z_2)Q(\xi_1, \bar\xi_1)...Q^+(\xi_{2N},
\bar\xi_{2N})\rangle \nonumber\\
&&=
\frac{1}{2N^2}\sum_j \frac{\sigma_j}{z_1 - \xi_j}\langle
C(z_2)Q(\xi_1, \bar\xi_1)...Q^+(\xi_{2N}, \bar\xi_{2N})\rangle -
\frac{1}{2N^2z_{12}^2}\langle
Q(\xi_1, \bar\xi_1)...Q^+(\xi_{2N}, \bar\xi_{2N})\rangle \label{wardc}
\end{eqnarray}
where $\sigma = 1$ for $Q$ and $-1$ for $Q^+$. Notice that the
operator $D$ does not appear in the right hand side of this identity.
This Ward identity is an important one since it, together with Eq.(\ref{CD})
establishes an isomorphism between the representations of
the Virasoro algebra and the algebra of the conserved current
$C$.

 Now let us study the fusion of $Q$ with itself. For this end it is
more convenient to rewrite the four-point correlation function
(\ref{four}) in terms of the solutions regular at $x \rightarrow
\infty$. For Eq.(\ref{equ}) these solutions are ($N \neq 2$)
\begin{eqnarray}
\tilde W_1^{(0)}(x) &&= (-x)^{-1/N}F(1/N,1/N,1 + 2/N; 1/x), \nonumber\\
 \tilde
W_2^{(0)}(x)
&&= (-x)^{-1/N}F(1 + 1/N,1/N,1 + 2/N;1/x)\nonumber\\
\tilde W_1^{(1)}(x) &&= (-x)^{1/N}F(- 1/N,- 1/N,1 - 2/N; 1/x), \nonumber\\
\tilde W_2^{(1)}(x)
&&= - (-x)^{1/N}F(1 - 1/N,- 1/N,1 - 2/N; 1/x)
\end{eqnarray}
These solutions have extremely simple monodromy properties:
\begin{eqnarray}
\tilde W_1^{(0)}(1 - x) &&= \tilde W_2^{(0)}(x), \:
 \tilde W_2^{(0)}(1 -
x) = \tilde W_1^{(0)}(x) \nonumber\\
 \tilde W_1^{(1)}(1 - x) &&= - \tilde
W_2^{(1)}(x), \:
 \tilde W_2^{(1)}(1 - x) = - \tilde W_1^{(1)}(x)
\end{eqnarray}
The crossing invariant form of the correlation function is
\begin{equation}
W(x, \bar x) = \alpha[\tilde W_1^{(0)}(x)\tilde W_2^{(0)}(\bar x) - k^2\tilde
W_1^{(1)}(x)\tilde W_2^{(1)}(\bar x) + (x \rightarrow \bar x)] \label{infty}
\end{equation}
where
\[
k = \frac{\Gamma(1 + 2/N)\Gamma^2(- 1/N)}{\Gamma(1 -
2/N)\Gamma^2(1/N)}
\]
The coefficient $\alpha$ whose numerical value we do not provide
should be  choosen to match Eq.(\ref{infty}) to
 the correlation function (\ref{zero})
regular at $x = 0$.
Let us
consider  the limit $z_{31} = \epsilon \rightarrow 0$ we have
\begin{equation}
G(1, 2;1 + 0, 2 +0) = 2\alpha|\epsilon|^{-4/N^2}[|z/\epsilon|^{4(N -
2)/N^2} -
 k^2|z/\epsilon|^{- 4(N + 2)/N^2}] + ...
\end{equation}
This expansion  is valid only for $N \neq 2$. In this case it
corresponds to the standard operator product expansion:
\begin{equation}
Q(1)Q(2) = C_1^{1/2}|z_{12}|^{-4\Delta + 2\Delta_A}O_A(2) +
C_2^{1/2}|z_{12}|^{-4\Delta + 2\Delta_S}O_S(2) + ...
\end{equation}
where $C_1, ~C_2$ are numerical coefficients and $O_A$ and $O_S$ are
operators
 from the asymmetric and the
symmetric representations whose Young tableaus are $(1,1,0, ...)$ and
$(2,0,...)$ respectively. Their conformal dimensions are given by
Eq.(\ref{dims}):
\begin{eqnarray}
\Delta_A = \frac{2 - N}{N^2}, \: \Delta_S = \frac{2 + N}{N^2} \label{eq:dim}
\end{eqnarray}
which  reproduces the  result obtained in the previous publications \cite{tsv}
and \cite{wen}.

At $N = 2$ the dimension of the antisymmetric  operator vanishes. Now
we have a situation where there are three operators with zero
conformal dimension - the unity, the adjoint operator and the operator
in the antisymmetric representation. This situation will be discussed
elsewhere.

\section{Deformation by the logarithmic  operators}

 The conventional WZNW model remains an integrable theory  even if one
changes the coefficient in front of the
Tr$(\partial_{\mu}g^+\partial_{\mu}g)$-term in the action
(4). According to \cite{knizh}, such  perturbation is equivalent to  the
$J_{-1}\bar
J_{-1}\Phi^{ab}$-operator (recall that $\Phi^{ab}$ is the primary
field in the adjoint representation). The
corresponding beta
function is
\begin{equation}
\beta(\gamma) = \frac{2c_v}{c_v + N}\gamma
\end{equation}
where $\gamma$ is the deviation of the coupling constant from its
critical value.
In our case $c_v = r \rightarrow 0$ and the beta function apparently
vanishes. This means that the perturbation becomes marginal and
we have to reconsider the terms of higher order in $\gamma$.
Despite the  fact that $\Phi^{ab}$ does not appear now in
OPE, its decendants, that is the logarithmic
operators $D, ~\bar D$  do appear.
We suggest that the
 change in the coupling constant of the WZNW model (4) is
associated with the perturbation by the marginal operator
$\gamma\bar{D}D$. We warn the reader not to confuse this perturbation with
a change of the disorder strength ${\cal A}$ which is truely
 irrelevant, leading to a
change of the cut-off $M$. One physical mechanism of a marginal  deformation
 away from criticality  in the  model (4) was described in
\cite{ners}  (see Chapter 7).  We conjecture that this
 deformation is  generated by  the $\gamma\bar{D}D$-perturbation.

 In the case of a deformation $\gamma \int \mbox{d}^2 z O(z, \bar{z})$
 caused by a  usual marginal
operator $O$ one has two possibilities depending on the operator
product expansion $$O(z, \bar{z}) O(0) = f \frac{O(0)}{|z|^2} + ...$$
The first one   is when $f =0$, i.e.  the  OPE of $O(z, \bar{z}) O(0)$
 does not
 contain the operator $O$ itself.
In this case   this  operator is truly marginal
 and one has  the continuous family of
 conformal field theories parametrized  by the deformation parameter
  (coupling constant). The
 anomalous dimensions $\Delta$ depend  on this  parameter. In the
model (1) this situation is realized when one introduces an Abelian
disorder (see \cite{wen}).
  In the opposite case, when $f \neq 0$, i.e.   the  OPE of $O(z) O(0)$
 contains  $O$ itself, there is a renormalization group (RG)
flow of the coupling constant
$$ \frac{\mbox{d}\gamma}{\mbox{d} \ln \Lambda} = f \gamma^2  +..$$
  which means that the theory actually depends on the scale $\Lambda$.

 Let us now study the same problem in a case where the theory is
deformed
 by the operator  $\bar{D}D$, which is truly marginal, because
  the OPE of $D(z) D(0)$ does not contain the operator $D$ itself
 (see Eq.(\ref{DD}).
 In this case we shall  calculate  the correlation function
 \begin{eqnarray}
G(z; \gamma) &&=
\langle A(z) B(0) \exp(\gamma \int \mbox{d}^2 x \bar{D}D(x)\rangle
\nonumber \\
&&= \sum_{n} \frac{\gamma^n}{n!}\int \langle A(z) B(0)~
\bar{D}D(x_1)....\bar{D}D(x_n)\rangle \mbox{d}^2 x_1......\mbox{d}^2 x_n
\end{eqnarray}
where $A$ and $B$ are some operators ($Q$ and $Q^{+}$, for example)
 with the correlation function
 $$
G(z; 0) =  \langle A(z) B(0)\rangle
$$
Using the OPE
\begin{equation}
\bar{D}D(x)~ A(y) = a \frac{\ln^2 |x-y|^2}{|x-y|^2} A(y),
{}~~~
\bar{D}D(x)~ B(y) = b \frac{\ln^2 |x-y|^2}{|x-y|^2} B(y)
\end{equation}
one  can  find the first  order in $\gamma$ correction to the
 correlation function which will be
\begin{eqnarray}
\gamma \int \langle &&A(z) B(0)~
\bar{D}D(x )\rangle \mbox{d}^2 x = \\ a&& \gamma
\langle A(z) B(0)\rangle \int \mbox{d}^2 x \frac{\ln^2
|x-z|^2}{|x-z|^2}
 ~ +  b \gamma
\langle A(z) B(0)\rangle \int \mbox{d}^2 x \frac{\ln^2
|x|^2}{|x|^2} \nonumber
\end{eqnarray}
 where in both integrals we integrate over $x$ between $0$ and $z$.
 Then it is easy to find the following  logarithmic correction:
$$
 \gamma \frac{ (a +b)}{3}~ \ln^3 |z|^2  G(z; 0)
$$
 Now one can consider the next order corrections and sum all of them
  using   the same methods as in the case of conventional  marginal
 operators (see \cite{Pokrovskii}). The result is
\begin{equation}
G(z; \gamma) = G(z; 0) \exp\left(\gamma\frac{a+b}{3}\ln^3 |z|^2\right)
\label{green}
\end{equation}
 which is different from  the case of a conventional  marginal operator when
 one has the first power of log in the exponent and not the third.
 The first power in the exponent introduces  the power
 factor
\begin{equation}
\exp\left((a+b) \gamma \ln |z|^2\right) = |z|^{2(a+b)\gamma}
\end{equation}
 corresponding to the change in the anomalous dimension
$\Delta_{A}(\lambda) = \Delta_{A}(0) - (a+b) \gamma$,  and the
 behaviour of the deformed correlation function is still power-like.
 The are no logarithmic corrections  after all. This is not true anymore
 with the logarithmic operator, when  the correlation function
cannot be written as a power at all.
  Thus  we see
 that the correlation functions for operators which have non-trivial
 OPE with the logarithmic operator $D$ (like our primary
 fields $Q$, for example) will have  logarithmic corrections in the
 deformed theory - even in the absence of the RG flow.

 Let $A = Q$ and $B = Q^+$, then, according to Eq.(\ref{cdq}), $a = b
= N^{-4}$. At $\gamma < 0$ the correlation function (\ref{green}) decays faster
then
any power. At $\gamma >  0$ it increases faster than any power. In this
case the approximation leading to Eq.(\ref{green}) breaks down when
the correlation function begins to increase, that is at
\begin{equation}
|z| \sim M^{-1}\exp[\frac{N}{4\sqrt\gamma}]
\end{equation}
We speculate that for $\gamma > 0$  the symmetry is broken and the finite
density of states at $\epsilon = 0$ is formed. This probably explains
the finite DOS obtained numerically in disordered d-wave
superconductors by Wheatley\cite{joe}.

\section{Probability distribution of local DOS}

 Now we shall calculate the distribution function of local densities
of states. We can do it for a system of a finite size $L$. From
Eq.(\ref{rho}) we know that in the case of zero frequency we have
\begin{equation}
\langle \rho^n(x)\rangle = M^n\langle [\mbox{Tr}(Q + Q^+)]^n\rangle
\sim L^{- 2\Delta_n} \label{cor}
\end{equation}
The latter equality is valid in  the leading
order in $1/L$; $\Delta_n$ is given by Eq.(\ref{deltas})
 being the smallest conformal dimension
in the operator
product of n operators Tr$\langle(Q +
Q^+)\rangle$. For $N > 2$ $\Delta_n$'s are  negative for $n > 1$.

 Let us imagine now that the result (\ref{cor}) comes from a local
distribution function of $\rho$:
\begin{equation}
\langle \rho^n(x)\rangle = \int_0^{\infty}P(\rho)\rho^n \mbox{d}\rho =
A_n\exp\{2\ln L[\frac{(N - 1)n^2}{2N^2} - \frac{n}{2N}]\}
\end{equation}
where $A_n$ may contain powers of  $\ln L$. The distribution function
which
reproduces this result is the famous log-normal  distribution which
is considered as a
characteristic feature of disordered systems \cite{lerner},
\cite{raz}, \cite{falko}:
\begin{eqnarray}
P(\rho) &&= D(\rho)\exp\left[ - \frac{1}{\ln L^{\eta}}\ln^2(\rho
L^{\zeta})\right], \label{lognorm}\\
 \zeta &&= \frac{1}{N}(3 - 2/N), \: \eta = \frac{4(N - 1)}{N^2}
\end{eqnarray}
where $D(\rho)$ is a smooth function of $\ln\rho$ which we cannot
determine.

 From the fact that the frequency scales as $L^{- 2 + 2\Delta_1}$ we
can conjecture that the distribution function of $\rho(\omega)$ is
given by
\begin{eqnarray}
P(\rho_{\omega}) =
 D(\rho_{\omega})\exp\left[ - \frac{1}{\ln (1/\omega^{\gamma})}
\ln^2(\rho_{\omega}
\omega^{- \beta})\right], \nonumber\\
 \beta = \frac{3N - 2}{2N^2 - 1}, \gamma = \frac{4(N - 1)}{2N^2 - 1}
\end{eqnarray}

The authors of Ref.\cite{wen} have discussed this distribution without
writing it down explicitly.

 It is worth remarking that the log-normal distribution also appears
in the Liouville theory (27). Indeed, according to Eq.(\ref{Eq}), the
conformal dimension of the operator $\exp(n\alpha\phi)$ is given by
$\frac{1}{2}n\alpha(Q - n\alpha)$, i.e. it has the same quadratic
$n$-dependence as $\Delta_n$ (\ref{deltas}). Repeating the previous arguments
we obtain for the vertex
operator $V = \exp(\alpha\phi)$ the log-normal distribution
(\ref{lognorm}) with $\eta = 4\alpha^2, ~ \zeta = \alpha Q(3 - 2
\alpha Q)$. Since the logarithmic operators appear in this theory only
when $Q = 2\alpha$, we conclude that their presence is not directly
related to multifractality of the target space.

\section{Conclusions}
 In this paper we have demonstrated that in the general class of
nonunitary  critical
models a new phenomenon takes place - the emergence of
 logarithmic operators associated with a special hidden continuous symmetry.
The presence of this symmetry is intimately related to the fact that
the order parameter of our model - the local density of states at
$\epsilon = 0$ - does
not acquire a non-zero average.  Since above mentioned
features appear also in quantum gravity \cite{bk}, we
anticipate  a connection between quantum gravity and critical models with
disorder \cite{future}.  Our expectations are supported by
the recently discovered  similarities  between the
conventional localization theory and  the Liouville theory
\cite{falko}, \cite{khm}.

 The physical meaning of the hidden symmetry in
 models with disorder as well as in 2d gravity, remains obscure.  It
is clear,  however, that this symmetry
should routinely appear
in critical non-unitary theories where the Hamiltonian
cannot be diagonalized (in our paper
this fact is expressed  in Eq.(\ref{example}).

  It follows also from our work that, at least for the model in
question, the replica approach is equivalent to the supersymmetric
one. This is a pleasant fact.

 As we have said above, Eqs.(\ref{CD}) and (\ref{wardc}) enable one in
principle  to
reformulate the theory in terms of representations of the current
algebra of the conserved current $C(z)$. With this task being accomplished
one can abanbon  replicas  and treat the theory
axiomatically as it is customary, for instance, in the theory of
the standard WZNW model. At present this remains the
biggest challenge.

\acknowledgments
The authors  are grateful to  B. Altshuler, J. Chalker, K. Efetov, V. Fal'ko,
A. Nersesyan,  D. Khmelnitskii, N. Mavromatos,
B. Muzykantskii, R. Stinchcombe and J. Wheater for  valuable and inspirational
discussions. Our special thanks are  to I. Lerner, who acquainted us
with the results obtained in the conventional localization theory.
One of us (I.I.K) would like also to thank
 A. Bilal and V. Gurarie for numereous interesting discussions about
 the logarithmic operators.

{\bf Appendix}

 In this Appendix we write down the relationship between conformal
blocks in the replica and the supersymmetric (SUSY) representations. According
to Ref.\cite{wen}, the correlation functions in SUSY representation
are products of correlation functions of the Gaussian model and the SU$_k$(N)
WZNW theory with $k = - 2N$. According to Ref.\cite{knizh}, the
conformal blocks of the latter model are
\begin{eqnarray}
W_1^{(0)}(x) &&= (1 - x)^{-1}F(1/N, - 1/N, 2; x)\nonumber\\
W_2^{(0)}(x) &&= - \frac{1}{2N}x(1 - x)^{-1}F(1 + 1/N, 1 - 1/N, 3; x)
\end{eqnarray}
The second solutions contain logarithms. The relationship between two
representations is established by the identity
\begin{eqnarray}
x(1 - 1/N^2)F(1 + 1/N, 1 - 1/N, 3; x) +  2(1 - x)F(1/N, - 1/N, 2; x)
\nonumber \\ =
(1 - xN^2)F(1/N, - 1/N, 1;x)
\end{eqnarray}
 using which one can   write the expression for four-point
function (\ref{four1}) in terms of conformal blocks either of replica
  or supersymmetric models.


\begin{references}
\bibitem{ners} A. A. Nersesyan, A. M. Tsvelik and F. Wenger, Phys. Rev.
Lett. {\bf 72}, 2628 (1994); Nucl. Phys. B.{\bf 438}, 561 (1995).
\bibitem {tsv} A. M. Tsvelik, Phys. Rev. B{\bf 51}, 9449
(1995).
\bibitem{wen} C. Mudry, C. Chamon, X.-G. Wen, unpublished.
\bibitem{gade} R. Gade, private communication.
\bibitem{efetov} K. B. Efetov, Adv. Phys. {\bf 32}, 53 (1983).
\bibitem{remark} This statement is usually called the Mermin-Wagner
theorem. In non-unitary theories which admit operators with
non-positive scaling dimensions such  operators may acquire non-zero
averages. However,  this does not concern the operator $Q$ whose scaling
dimension is positive.
\bibitem{gorkov} L. P. Gor'kov and P. A. Kalugin, Pis'ma v ZhETP{\bf
41}, 208 (1985)
[JETP Lett. {\bf 41}253 (1985)].
\bibitem{rs} L. Rozansky and H. Saleur, Nucl. Phys. B{\bf 376}, 461 (1992).
\bibitem{gurarie} V. Gurarie, Nucl. Phys. B{\bf 410}, 535 (1993);
hep-th/9303160.
\bibitem{bk} A. Bilal and I.I. Kogan, PUPT-1482, hep-th@xxx/9407151
(unpublished); Nucl. Phys. B{\bf 449}, 569 (1995); hep-th@xxx/9503209.
\bibitem{polch} J. Polchinski,  Nucl. Phys. B{\bf 346}, 253 (1990);
Nucl. Phys. B{\bf 357}, 241 (1991).
\bibitem{knizh} V.G. Knizhnik and A. B. Zamolodchikov, Nucl. Phys.
{\bf B247}, 83 (1984).
\bibitem{abram} {\it Handbook of Mathematical Functions}, Ed. by M.
Abramovitz and I. A. Stegun, ed. by National Bureau of Standards, (1964).
\bibitem{Pokrovskii} V. L. Pokrovskii and B. Patashinskii, {\it
Fluctuations Theory of  Phase Transitions} ~~ (in Russian),
 Moscow, 1975.  The method  used in this book is due to \\
 A. M. Polyakov, Sov. Phys. ZhETF {\bf 63}, 24 (1972).
\bibitem{joe}. J. Wheatley, preprint.
\bibitem{lerner} B. Altshuler, V. Kravtsov and I. Lerner,
Sov. Phys. JETP Lett. {\bf 43}, 441; Sov. Phys. JETP {\bf 64}, 1352
(1986); Phys. Lett. A{\bf 134}, 488 (1989).
\bibitem{raz} I. Lerner, Phys. Lett. A{\bf 133}, 253 (1988).
\bibitem{future} I. I. Kogan and A.M. Tsvelik, work in progress.
\bibitem{falko} K. B. Efetov and V. I.  Fal'ko, preprint.
\bibitem{khm} D. E. Khmelnitskii and B. A. Muzykantskii, preprint.

\end{references}
\end{document}